\documentclass[journal]{IEEEtran}

\usepackage{pgfplots}
\pgfplotsset{compat=newest}
\usepackage{subfigure}
\usepackage[cmex10]{amsmath}
\usepackage{amssymb}
\usepackage{bbold}
\usepackage{dsfont}
\usepackage{bm}
\usepackage{cite}
\usepackage{algorithmic}
\usepackage{algorithm}
\usepackage{tikz}
\usetikzlibrary{positioning,shapes,shapes.geometric,calc,chains,shapes.multipart,arrows}
\graphicspath{{fig/}}

\tikzset{
	rubberduck/.style={
		shape=isosceles triangle,
		fill=red!30,
		minimum height=18pt,
		minimum width=12pt,
		shape border rotate=#1,
		isosceles triangle stretches,
		inner sep=0pt,
	},
	ap/.style={rubberduck=+90},
	ducky/.style={rubberduck=-90},
	queue/.style={
		rectangle split,
		minimum width=2em,
		rectangle split parts=5,
		draw,
		anchor=south,
	}
}



\newcommand*\dif{\mathop{}\!\mathrm{d}}

\newcommand{\Kcal}{K}
\newcommand{\Lcal}{L}
\newcommand{\Ncal}{N}
\newcommand{\Pcal}{P}

\newcommand{\Rcal}{R}
\newcommand{\Scal}{S}

\renewcommand{\P}{\bm{P}}

\newcommand{\q}{\bm{q}}
\renewcommand{\r}{\bm{r}}
\newcommand{\w}{\bm{w}}
\newcommand{\x}{\bm{x}}
\newcommand{\y}{\bm{y}}
\newcommand{\p}{\bm{p}}

\newcommand{\z}{\bm{z}}
\renewcommand{\v}{\bm{v}}
\renewcommand{\u}{\bm{u}}
\renewcommand{\t}{\bm{t}}
\renewcommand{\w}{\bm{w}}

\ifCLASSINFOpdf
\else
\fi
\begin{document}
%
\title{A Centralized Metropolitan-Scale Radio Resource Management Scheme}
%
%
%

\author{Zhiyi~Zhou~and~Dongning~Guo
\thanks{The authors are with the Department of Electrical Engineering and Computer Science, Northwestern University, Evanston, IL.}}

%
%

\markboth{Zhou, Guo: A Centralized Metropolitan-Scale Radio Resource Management Scheme}%
{Shell \MakeLowercase{\textit{et al.}}: Bare Demo of IEEEtran.cls for IEEE Journals}
%



\maketitle

\begin{abstract}
This work studies centralized radio resource management in metropolitan area networks with a very large number of access points and user devices. A central controller collects time-averaged traffic and channel conditions from all access points and coordinates spectrum allocation, user association, and power control throughout the network on an appropriate timescale. The timescale is conceived to be seconds in today's networks, and it is likely to become faster in the future. The coordination problem in each time epoch is formulated as a network utility maximization problem, where any subset of access points may use any parts of the spectrum to serve any subsets of devices. It is proved that the network utility can be maximized by an extremely sparse spectrum allocation. By exploiting this sparsity, an efficient iterative algorithm with guaranteed convergence is developed, each iteration of which is performed in closed form. The proposed centralized optimization framework can incorporate a broad class of utility functions that account for weighted sum rates, average packet delay, and/or energy consumption, along with very general constraints on transmission powers. Numerical results demonstrate the feasibility of the algorithm for networks with up to 1,000 access points and several thousand devices. Moreover, the proposed scheme yields significantly improved throughput region and average packet delay comparing with several well-known competing schemes.
\end{abstract}


%
\IEEEpeerreviewmaketitle

\section{Introduction}
\label{intro}
%
%
%
%
\IEEEPARstart{W}{ireless} systems have emerged as a ubiquitous part of modern data communication networks. In order to meet the ever increasing demand for wireless data services, a large number of access points (APs) of different form factors and capabilities are being deployed to improve coverage and capacity for homes, businesses, and public spaces. These APs may be densely deployed and may utilize a wide range of spectrum resources, including millimeter wave bands and unlicensed spectrum resources.

Efficient resource allocation (e.g., spectrum allocation, power control, link scheduling, routing, and congestion control) is crucial to achieving high performance and providing satisfactory quality-of-service (QoS). Conventional spectrum allocation schemes include full spectrum reuse, partial frequency reuse, and fractional frequency reuse \cite{Resource:Zhang,Radio:Xue,Frac:Saquib}. These schemes have very limited spectrum agility and are generally not adaptive to traffic conditions. In today's cellular and WiFi networks, devices typically associate with the base station from which they receive the strongest signal. Power control is another way of mitigating inter-cell interference as well as saving energy. Instead of letting each AP always transmit at full power, some inter-cell interference cancellation techniques, such as the almost blank subframe (ABS) control, have been proposed in LTE and LTE-A. However, the performance of such kind of distributed algorithms is far from optimal especially for large networks with irregular AP placements. 

Resource allocation in wireless networks has been extensively studied over the last few decades. Spectrum allocation schemes are basically designed from both economic perspectives (e.g., \cite{The:Chen,rebato2016hybrid}) and technical perspectives (e.g., \cite{zhou2017cell,Energy:Kuang,Kuang:Optimal,Stolyar:Self,Chang:Multicell,Ali:Dynamic,Madan:Cell,Liao:Base,Etkin:Spectrum,Joint:Huang,Joint:Fool,Energy:Lim,Distributed:Shen,Algorithms:Deb,Licensed:Zhou,User:Ye,An:Ao,Traffic:Zhuang,Energy:Zhuang,Large:Zhuang,teng2015Dual}). However, it remains elusive to achieve the maximum spectrum flexibility and utilization. As for the user association problem, schemes based on game theory or optimization have also been well explored (e.g., \cite{Joint:Huang,Joint:Fool}). It is typically hard to obtain a local optimum since the formulated problems often include integer variables and nonconvex constraints. Similar to the user association problem, power control is also known to be a hard non-convex optimization problem due to the complicated interference coupling between links. Numerous distributed power control algorithms (e.g., \cite{Joint:Elbatt,Dis:Kub,Eff:Sar}) have been proposed to make proper power assignments for large-scale networks with substantially degraded system performance. 

In this paper, we address the radio resource management problem by proposing a centralized optimization-based framework. A distinguishing feature here is that we propose to use a \textit{central controller} or \textit{cloud} to coordinate a large network with many thousand APs. This architecture can fully harness the power of cloud computing, big data, and large-scale optimization methods. Specifically, APs measure/estimate traffic and channel conditions and send the cloud regular updates. The cloud periodically solves a large-scale optimization problem for the whole network and returns the resulting allocation plan or recommendations to the APs. The cloud may also exchange information with peer network operators in case of shared spectrum). 

The timescale of an allocation period is currently conceived to be on the order of seconds to allow the cloud sufficient time to solve a large-scale allocation problem using today's technologies. This timescale is fast enough to track demand shifts and large-scale fading.
On this timescale, resource allocation is likely to be in a relatively coarse granularity (e.g., blocks of subcarriers). The cloud's communication overhead is small. For example, if the cloud collects one million 8-bit parameters from one thousand cells every second, the overhead is about 1 KB/s per cell or 1 MB/s in total. As technologies improve over time, the timescale of centralized resource allocation is expected to become faster and the allocation is expected to be in finer granularities. Ultimately, the timescale is limited by the latency for collecting information from all APs via backhaul links, which maybe as short as the duration of a frame.

The proposed scheme jointly solves the problems of spectrum allocation, user association, and power control for very large networks. An efficient algorithm with low complexity is proposed to obtain a locally-optimal allocation solution in a timely manner. The proposed algorithm complements our previous work \cite{zhou2017cell} which considered spectrum allocation and user association but did not allow power control. One key distinction of the problem formulation in this paper is that, instead of assuming all APs always transmit using full power, we allow each AP to apply an arbitrary power spectral density over the entire band.

The main contributions of this paper are summarized as follows:
\begin{itemize}
	\item To the best of our knowledge, we are the first to propose to carry out centralized joint spectrum allocation, user association, and power allocation in metropolitan-scale networks with thousands of APs and user devices.
	\item We first formulate a problem of joint continuous spectrum allocation, power control, and user association. We then reduce this infinite size problem to an equivalent finite-dimensional optimization problem which is amenable to a very efficient algorithm. 
	\item We propose an efficient algorithm with guaranteed convergence to solve the problem numerically. Each iteration is performed in closed form, which makes centralize resource allocation practical for very large networks.
	\item Through simulations, we demonstrate that the proposed resource allocation scheme yields significantly improved throughput region and average packet delay comparing with several well-known competing schemes.
	\item We generalize the problem formulation and algorithm to accommodate a much broader set of utility functions, spectral efficiencies, and power constraints than in most other studies.
	\item We develop a rigorous proof of the sufficiency of using piecewise constant power allocation to achieve the global optimum of the resource allocation problem under the most general set of conditions in the literature.
\end{itemize} 

The remainder of this paper is organized as follows. The system model is introduced in Sec.~\ref{sys model}. The problem formulation is given in Sec.~\ref{problem formulation}. An efficient algorithm is proposed in Sec.~\ref{sec:alg}. A significantly generalized resource management problem is studied in Sec.~\ref{sec:general}. Simulation results are presented in Sec.~\ref{numerical}. Concluding remarks are given in Sec.~\ref{conclusion}. 
{Most} technical proofs are relegated to the appendices.

\section{System Model}
\label{sys model}
We consider the downlink of a single-input single-output (SISO) network of $n$ APs, $k$ user devices,\footnote{Referred to as user equipment in LTE standards.} and a central controller. The APs and the controller are connected through a wired network including backhauls. The controller continuously collects all traffic and channel/interference information from the APs. At the disposal of the controller is a radio frequency band of $W$ Hertz. The task of the controller is to allocate the radio spectrum and power resources to all AP-device links in order to maximize a certain long-term network utility. 

The central controller performs resource allocation on a moderate timescale, which is conceived to be once every several seconds in today's networks. This allows ample time for information exchange (including channel state information feedback) and to carry out a large-scale optimization at the controller. In addition, since this timescale is much longer than the channel coherence time, the average channel conditions can be represented using path loss and the statistics of small scale fading. On this timescale, it is fair to assume that large blocks of the spectrum are homogeneous in the sense that all hertz are equally valuable and interchangeable. This spectral homogeneity assumption can be relaxed as the proposed formulation and solution can be easily generalized to treat multiple bands of different characteristics\cite{Licensed:Zhou}.

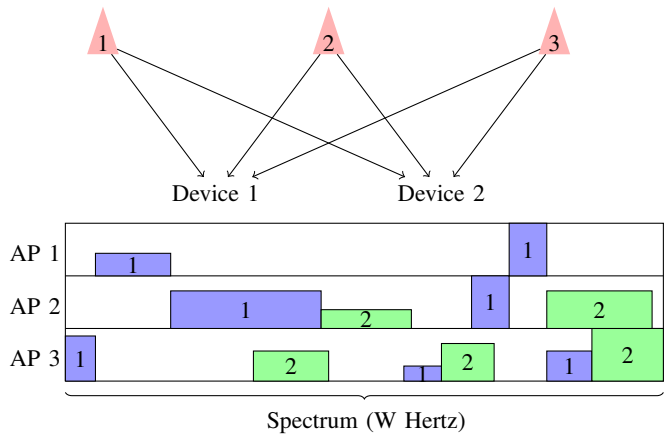
\begin{figure}
	\centering
	\begin{tikzpicture}[x=1cm,y=1cm,node distance=0 cm,outer sep=0pt]
	\small
	\tikzstyle{specb}=[rectangle,draw,anchor=south west,text centered,minimum height=7mm,fill=blue!10]
	\tikzstyle{specg}=[rectangle,draw,anchor=south west,text centered,minimum height=7mm]
	
	\draw (0,0) -- (0,2.1) -- (7.95,2.1) -- (7.95,0) -- (0,0);
	\draw (0,0.7) -- (7.95,0.7);
	\draw (0,1.4) -- (7.95,1.4);
	\node at (-0.4,0.3) {AP 3};
	\node at (-0.4,1) {AP 2};
	\node at (-0.4,1.7) {AP 1};
	\node at (4,-.55) {Spectrum (W Hertz)};
	
	\filldraw[fill=blue!40!white, draw=black] (0,0) rectangle (0.4,0.6);
	\node at (0.2,0.3) {1};
	\filldraw[fill=green!40!white, draw=black] (2.5,0) rectangle (3.5,0.4);
	\node at (3,0.2) {2};
	\filldraw[fill=blue!40!white, draw=black] (4.5,0) rectangle (5,0.2);
	\node at (4.75,0.1) {1};
	\filldraw[fill=green!40!white, draw=black] (5,0) rectangle (5.7,0.5);
	\node at (5.35,0.25) {2};
	\filldraw[fill=blue!40!white, draw=black] (6.4,0) rectangle (7,0.4);
	\node at (6.7,0.2) {1};
	\filldraw[fill=green!40!white, draw=black] (7,0) rectangle (7.95,0.7);
	\node at (7.475,0.35) {2};
	
	\filldraw[fill=blue!40!white, draw=black] (1.4,0.7) rectangle (3.4,1.2);
	\node at (2.4,0.95) {1};
	\filldraw[fill=green!40!white, draw=black] (3.4,0.7) rectangle (4.6,0.95);
	\node at (4,0.825) {2};
	\filldraw[fill=blue!40!white, draw=black] (5.4,0.7) rectangle (5.9,1.4);
	\node at (5.65,1.05) {1};
	\filldraw[fill=green!40!white, draw=black] (6.4,0.7) rectangle (7.8,1.2);
	\node at (7.1,0.95) {2};
	
	\filldraw[fill=blue!40!white, draw=black] (0.4,1.4) rectangle (1.4,1.7);
	\node at (0.9,1.55) {1};
	\filldraw[fill=blue!40!white, draw=black] (5.9,1.4) rectangle (6.4,2.1);
	\node at (6.15,1.75) {1};
	
	\draw[decorate, decoration={brace,mirror}, black] (0,-.15) -- (7.95,-.15);
	
	\draw (.5,4.5) node[ap] (apone) {1};
	\draw (3.5,4.5) node[ap] (aptwo) {2};
	\draw (6.5,4.5) node[ap] (apthree) {3};
	\node (ueone) at (2,2.5) {Device 1};
	\node (uetwo) at (5,2.5) {Device 2};
	\draw [->] (apone) -- (ueone);
	\draw [->] (aptwo) -- (ueone);
	\draw [->] (aptwo) -- (uetwo);
	\draw [->] (apthree) -- (uetwo);
	\draw [->] (apthree) -- (ueone);
	\draw [->] (apone) -- (uetwo);
	\end{tikzpicture}
	\caption{Illustration of an example of resource allocation in a 3-AP 2-device network.}
	\label{fig1}
\end{figure}
Denote the set of AP indexes by $\Ncal=\{1,\dots,n\}$ and the set of device indexes by $\Kcal=\{1,\dots,k\}$. We consider highly flexible resource allocation, in which an AP may serve any device at parts of the frequency band using any power (subject to power constraint), and a device may be served by any sets of APs.\footnote{In Sec.~\ref{sec:fast}, we introduce a device centric model which only includes viable links in the computation.} Fig.~\ref{fig1} gives an example of resource allocation for a small network with three APs and two devices. In Fig.~\ref{fig1}, the rectangles in each row are the spectrum segments used by the corresponding AP to serve the device whose index is marked on that spectrum segment. For each rectangle, its width represents the occupied bandwidth and its height represents the transmit power level. Fig.~\ref{fig1} also determines user association as AP 1 only serves device 1 while AP 2 and AP 3 serve both devices. Moreover, some links are assigned spectrum exclusively (the link from AP 1 to device 1), resulting in relatively high spectral efficiency. Some other links suffer interference from other APs operating on the same part of the spectrum, treating these interference as noise leads to relatively low spectral efficiency.\footnote{While treating interference as noise is the most practical way, cooperative transmissions have recently been incorporated in a similar model \cite{Jing:Co}.} 

The allocation problem is to determine the power spectral density functions (PSDs) of all AP-device links over the entire spectrum in order to maximize the network utility. Denote the PSD of the link from AP $i$ to device $j$ as a function $p_{i\to j}(f), f\in[0,W]$ and denote the PSDs of all links by $\p(f) = (p_{i\rightarrow j}(f))_{i\in\Ncal,j\in\Kcal}$. Clearly, $(\p(f),~f\in[0,W])$ completely characterizes power spectrum allocation as well as user association in the network. Choosing a subset of AP-device links to transmit over a certain part of the spectrum is equivalent to allocating positive transmit power to these AP-device links and zero power to others over that part of the spectrum. For instance, the allocation illustrated by Fig.~\ref{fig1} corresponds to an array of piecewise constant PSDs.

For now we assume the following peak power constraint due to physical and regulatory reasons:
\begin{align}
0\leq p_{i\rightarrow j}(f)\leq P_{\text{max}}, \quad \forall i\in \Ncal, j\in\Kcal, f\in[0,W].\label{eee}
\end{align}
We also assume that if an AP serves multiple devices, it uses orthogonal spectrum to do so. Mathematically,
\begin{align}
\sum\limits_{j\in\Kcal}{|p_{i\rightarrow j}(f)|}_0 \leq 1, \quad  \forall i\in \Ncal, f\in[0,W],
\label{eq1}
\end{align} 
where ${|x|}_0 = 1$ if $x \neq 0$, and ${|x|}_0 = 0$ if $x = 0$.

Constraints (\ref{eee}) and (\ref{eq1}) are intrinsic to most radio system designs. We adopt them for simplicity and practicality. In Sec.~\ref{sec:general} we replace those constraints to much more general forms, including total power constraint and broadcasting to multiple devices over the same slice of spectrum.

Define $g_{i\rightarrow j}$ as the average channel gain of link $i\rightarrow j$ which captures the effects of path loss and shadowing. These gains can be measured or calculated using the path loss model specified in the LTE standard \cite{3GPP:LTE}. Let $s_{i\rightarrow j}(\cdot): \mathbb{R}^{n\times k}\rightarrow \mathbb{R}$ represent a suitable spectral efficiency function which can either be calculated based on pathloss and other impairments or be measured over time. One popular choice is to use Shannon's formula to define: 
\begin{align}
s_{i\rightarrow j}(\p) = \log\left(1+\gamma_{i\rightarrow j}(\p)\right),\quad \forall i\in \Ncal, j\in\Kcal, \label{eq3}
\end{align}
where the signal-to-interference-and-noise ratio of link $i\rightarrow j$ is
\begin{align}
\gamma_{i\rightarrow j}(\p)=\frac{p_{i\rightarrow j}g_{i\rightarrow j}}{n_0 + \sum\limits_{(l,q)\in(\Ncal\times\Kcal)\backslash(i,j)}g_{l\rightarrow j}p_{l\rightarrow q}}, \label{eqsinr}
\end{align}
where $\p={(p_{i\rightarrow j})}_{i\in\Ncal,j\in\Kcal}$ is a double array of numbers representing link powers, $n_0$ denotes the PSD of white Gaussian noise. Note that $(\Ncal\times\Kcal)\backslash(i,j)$ denotes the set of AP-device links except for the link $i\rightarrow j$. Hence, $\sum\limits_{(l,q)\in(\Ncal\times\Kcal)\backslash(i,j)}g_{l\rightarrow j}p_{l\rightarrow q}$ is the PSD of the total interference from all the APs.

Given $\p(f)$, the service rate to device $j$ contributed by AP $i$ can be calculated by integrating the spectral efficiency of link $i\rightarrow j$ over the entire spectrum: 
\begin{align}
r_{i\rightarrow j} = \int_{0}^{W} s_{i\rightarrow j}(\p(f)) \dif f,\quad \forall i\in \Ncal, j\in\Kcal.
\label{general}
\end{align}
Throughout this paper, all integrals are defined in Lebesgue's sense.\footnote{Specifically, let $\mu$ denote the Lebesgue measure on $E=[0,W]$. Then the Lebesgue integral of a measurable function $h$ on $E$ is $\int_{E}h \dif \mu$, which is also written as $\int_{0}^{W}h(f)\dif f$ in this paper for convenience.}

The total service rate of device $j$ denoted as $r_j$ can be calculated by summing the service rates contributed from all APs:
\begin{align}
r_{j}=\sum\limits_{i\in\Ncal}r_{i\rightarrow j}.
\label{eq4}
\end{align}

\section{Problem Formulation} 
\label{problem formulation}
Let the long-term utility be given by a function $u(\r)$ of the rate tuple $\r=(r_1,\cdots,r_k)$. The goal is to maximize the long-term utility by adapting the PSDs of all AP-device links. Collecting the constraints (\ref{eee})--(\ref{eq4}), we formulate the problem as P0:
\begin{alignat*}{2}
\mathop{\text{maximize}}\limits_{(r_j), (p_{i\rightarrow j}(f))}\ &u(\bm{r})\tag{P0a} \label{P1a}\\
\text{subject to} \ \ \
&r_{j}=\sum\limits_{i\in\Ncal}\int_{0}^{W} s_{i\rightarrow j}(\p(f)) \dif f, \ \forall j\in \Kcal \tag{P0b} \label{P1b} \\
&\sum\limits_{j\in\Kcal}{|p_{i\rightarrow j}(f)|}_0 \leq 1,  \forall i\in \Ncal, f\in[0,W]\tag{P0c} \label{P1c}\\
&0\leq p_{i\rightarrow j}(f)\leq P_{\text{max}}, \forall i\in \Ncal, j\in\Kcal, f\in[0,W]. \tag{P0d} \label{P1d} 
\end{alignat*}
The variables are $(r_j)$ and $(p_{i\rightarrow j}(f))$, also denoted as $\r$ and $\p(f)$. Equation (\ref{P1b}) requires that $\p(\cdot)$ is such that $s_{i\rightarrow j}(p(f))$ is Lebesgue integrable. Also, we will justify the use of ``maximize'' in (\ref{P1a}), i.e., that the maximum is actually achievable in Appendix~\ref{append_A}.

Most work on resource management uses the sum rate, the minimum device service rate (max-min fairness), and the sum log-rate (proportional fairness) as the utility function. The techniques developed here are applicable to all those and many other utility functions of interest. For concreteness, we consider the average packet delay \cite{Traffic:Zhuang} as the utility function, that is
\begin{align}
u(\r) = -\sum_{j\in \Kcal}\frac{\lambda_j}{(r_j-\lambda_j)^+},
\label{eq5}
\end{align}
where $\lambda_j$ represents the amount of traffic intended for device $j$. The choice of this utility function is justified as follows. Suppose all downlink traffic arrive at all APs with Poisson packet arrival rate and exponential packet length. Then the packets for all the devices form $k$ independent M/M/1 queues and (\ref{eq5}) is minus the derived average packet delay. The utility (\ref{eq5}) can be generalized to accommodate various kinds of distributions \cite{Licensed:Zhou}. In (\ref{eq5}), the extended real-valued function $1/x^+ = 1/x$ if $x>0$ and $1/x^+ = +\infty$ if $x\leq 0$. It is easy to see that $1/x^+$ is convex on $(-\infty,+\infty)$. If $r_j \leq \lambda_j$, the packet delay is infinite, i.e., the system becomes unstable. We remark that, although the sum rate is often used as the utility, delay is generally a more relevant system-level figure of merit on a moderate timescale, which allows resource allocation to adapt to traffic conditions. In particular, it does not starve links with poor channel conditions.

P0 is computationally infeasible due to two factors: (i) the functional variable $p_{i\rightarrow j}(f)$ on $[0,W]$ implies an infinite problem size; (ii) constraints (\ref{P1b}) and (\ref{P1c}) are nonconvex. In the next subsections, we will first reformulate P0 to a problem with a finite number of variables, and then propose an efficient algorithm to overcome the aforementioned difficulties.

\subsection{Finite-Size Problem Reformulation}
We first reveal important characteristics of an optimal solution to P0, which leads to a dramatically simplified reformulation of the resource allocation problem.

\textit{Definition 1}:
A power allocation $(p_{i\rightarrow j}(f))_{i\in\Ncal,j\in\Kcal}$ is said to be {\em $m$-piecewise constant} if the interval $[0,W]$ can be partitioned into $m$ sub-intervals such that, for every $i\in \Ncal$, $j\in\Kcal$, $p_{i\rightarrow j}(\cdot)$ is constant on everyone of those sub-intervals. 

\textit{Theorem 1}: 	
The optimal utility of P0 can be attained by a $(k+1)$-piecewise constant power allocation.	If $u(\r)$ is increasing in every dimension of $\r$, then a $k$-piecewise constant allocation suffices.

Theorem~1 is proved in Appendix~\ref{append_B}. Theorem~1 guarantees that any optimal rate vector $\r$ can be achieved by dividing the entire spectrum into $k$ intervals with bandwidths $\beta^1, \cdots, \beta^k$ on which all link PSDs are flat (assuming $u(\r)$ is increasing in $\r$). The key to the proof is that all hertz are interchangeable and the frequency resource is additive. Let $\Lcal = \{1,\cdots,k\}$ denote the set of indexes of these $k$ intervals.\footnote{Although the set $L$ is defined to be identical to the set $\Kcal$ here, we use a different notation to allow a flexible choice of L subsequently.} Power allocation over the $m$-th interval is represented by a vector $\p^m = (p_{1\rightarrow 1}^m, \cdots, p_{k\rightarrow n}^m)$. Therefore, P0 can reformulated as P1: 
\begin{alignat*}{2}
&\mathop{\text{maximize}}\limits_{(p^m_{i\rightarrow j}),(\beta^m),(r_j)}\ \ \ \  u(\r)\tag{P1a} \label{P2a}\\
&\text{subject to} \ \ \ \ \\
&r_j = \sum\limits_{m\in\Lcal}\beta^m\sum\limits_{i\in\Ncal}\log\left(1+\frac{p^m_{i\rightarrow j}g_{i\rightarrow j}}{n_0 + \sum\limits_{(l,q)\in(\Ncal\times\Kcal)\backslash(i,j)}g_{l\rightarrow j}p^m_{l\rightarrow q}}\right), \\
&\ \ \ \ \ \ \ \ \ \ \ \ \ \ \ \ \ \ \ \ \ \ \ \ \ \ \ \ \ \ \ \ \ \ \ \ \ \ \ \ \ \ \ \ \ \ \ \ \ \ \ \ \  \forall j\in\Kcal \tag{P1b} \label{P2b}\\
&\sum\limits_{j\in\Kcal}{|p^m_{i\rightarrow j}|}_0 \leq 1, \quad \forall i\in \Ncal, m\in\Lcal \tag{P1c} \label{P2c}\\
&\sum\limits_{m\in\Lcal}\beta^m = W,  \tag{P1d} \label{P2d}    \\
&0\leq p^m_{i\rightarrow j}\leq P_{\text{max}}, \quad \forall i\in \Ncal, j\in\Kcal, m\in\Lcal \tag{P1e} \label{P2e}\\
&\beta^m\geq 0, \quad \forall m\in\Lcal. \tag{P1f} \label{P2f} 
\end{alignat*}

In contrast to P0 with infinite dimensions, P1 has only $k + k^2n = O(k^2n)$ real-valued variables. For a large network consisting of thousands of APs and devices, the number of variables is still prohibitively large. In addition, the set of constraints (\ref{P2b}) and (\ref{P2c}) are nonconvex. Hence P1 is still difficult to solve. In the following subsections, we will reveal some important facts in order to solve a simple case of P1 where the utility function $u(\r)$ is an affine function of $\r$. The technique motivates the development of the highly scalable algorithm in Sec.~\ref{sec:alg}.

\subsection{Affine Utility Function}
\textit{Proposition 1}: Suppose the utility function $u(\r)$ is affine, i.e., it can be written as 
\begin{align}
u(\r) = d + \sum\limits_{j\in\Kcal}c_jr_j,
\label{eq77}
\end{align}
for some constants $d$ and $(c_j)_{j\in\Kcal}$. Then the maximum utility in P1 can be attained by letting each AP apply a flat PSD over the entire spectrum to serve a single device (or no device if the PSD is all zero).

\textit{Proof}: Let $\p^m$ represent the array ${(p^m_{i\rightarrow j})}_{i\in\Ncal, j\in\Kcal}$. Then P1 can be rewritten in the following equivalent form, referred to as P2:
\begin{alignat*}{2}
&\mathop{\text{maximize}}\limits_{(p^m_{i\rightarrow j}),(\beta^m)}\ \ \ \ d+\sum\limits_{m\in\Lcal}\beta^m\sum\limits_{j\in\Kcal}c_j\sum\limits_{i\in\Ncal}s_{i\rightarrow j}(\p^m) \tag{P2a} \label{P3a}\\
&\text{subject to} \ \ \ \
(\ref{P2c}), (\ref{P2d}), (\ref{P2e}), (\ref{P2f}).
\end{alignat*}
Let $(\p,\bm{\beta})$ be a feasible solution to P2. Define $m^*\in\Lcal$ as a maximizer of $\sum\limits_{j\in\Kcal}c_j\sum\limits_{i\in\Ncal}s_{i\rightarrow j}(\p^m)$. It is easy to see that for any fixed power allocation $(p^m_{i\rightarrow j})$, the utility is always maximized by letting $\beta^{m^*} = 1$ and $\beta^m=0$ for all $m\neq m^*$, where the constraints (\ref{P2c})-(\ref{P2f}) remain to hold. Hence letting each AP apply a flat power spectral density over the entire spectrum to serve no more than one device is optimal. \hfill $\blacksquare$

From now on we refer to one assignment of link powers ${(p_{i\rightarrow j})}_{i\in\Ncal, j\in\Kcal}$ as a (power) profile. Proposition 1 implies that when the utility function is affine (which can be regarded as the weighted sum rate), it is optimal to apply a single power profile over the entire spectrum, where each AP serves a single device. Therefore, for the affine utility function (\ref{eq77}), P1 (and P0) is reduced to the following P3:
\begin{alignat*}{2}
&\mathop{\text{maximize}}\limits_{(p_{i\rightarrow j})}\ \ \ \ \\
&\sum\limits_{j\in\Kcal}c_j\sum\limits_{i\in\Ncal}\log\left(1+\frac{p_{i\rightarrow j}g_{i\rightarrow j}}{n_0 + \sum\limits_{l\in\Ncal:l\neq i}g_{l\rightarrow j}\sum\limits_{q\in\Kcal}p_{l\rightarrow q}}\right) \tag{P3a} \label{P30a}\\
&\text{subject to} \quad \sum\limits_{j\in\Kcal}{|p_{i\rightarrow j}|}_0 \leq 1, \quad \forall i\in \Ncal \tag{P3b} \label{P3b}\\
&\ \ \ \ \ \ \ \ \ \ \ \ \ \ \ 0\leq p_{i\rightarrow j}\leq P_{\text{max}}, \quad \forall i\in \Ncal, j\in\Kcal. \tag{P3c} \label{P3c}
\end{alignat*}

\subsection{A Fast Algorithm in case of Affine Utility}\label{sec:fast}
In the following, we first reformulate P3 to remove the $\ell_0$ norm constraint (\ref{P3b}) and further reduce the number of variables, and then solve it using the quadratic transform technique proposed by Shen and Yu \cite{shen2017fractional,shen2017fractional2}. 

In a large network with many APs, a device can in general only be served by a small subset of nearby APs due to path loss. For every $j\in\Kcal$, let $\Ncal_j$ denote the set of APs whose received signal-to-noise ratios at device $j$ are above a certain threshold $\xi$, i.e.,
\begin{align}
\Ncal_j \triangleq \left\lbrace i\in\Ncal~|~p_i g_{i\to j}>\xi n_0\right\rbrace,\quad \forall j\in\Kcal. \label{candidate}
\end{align}
We define $\Ncal_j$ as the {\em neighborhood} of device $j$. Device $j$ treats all APs outside $\Ncal_j$ as stationary noise sources whose total PSD (including noise) is denoted as $n_j$. This can be arbitrarily precise as $\Ncal_j$ may include all APs except those received by device $j$ at well below the noise level. Then for each AP $i$, define the set of devices that AP $i$ can potentially serve as $\Kcal_i$, specifically,
\begin{align}
\Kcal_i  \triangleq \left\lbrace j\in\Kcal~\bigg|~i\in \Ncal_j\right\rbrace,\quad \forall i\in\Ncal. \label{c}
\end{align}
It is fair to assume the size of all $\Ncal_j$s and $\Kcal_i$s are upper bounded by a constant $\alpha$, i.e., $|\Kcal_i|, |\Ncal_j|\leq \alpha$.

Since Proposition 1 guarantees that each AP serves no more than one device, let variable $z_i$ denote the device served by AP $i$, where $z_i=0$ if AP $i$ serves no device. For convenience, we introduce additional parameters $c_0=0$ and $g_{i\rightarrow 0}=0$ for all $i\in\Ncal$. Then based on Proposition 1, P3 can be reformulated as P4:
\begin{alignat*}{2}
\mathop{\text{maximize}}\limits_{\z,\p}\ \ \ \ &\sum\limits_{i\in\Ncal}c_{z_i}\log(1+\frac{p_{i}g_{i\rightarrow z_i}}{n_{z_i} + \sum\limits_{l\in\Ncal_{z_i}:l\neq i}p_{l}g_{l\rightarrow z_i}}) \tag{P4a} \label{P4a}\\
\text{subject to} \ \ \ \
&0\leq p_{i}\leq P_{\text{max}}, \quad \forall i\in \Ncal \tag{P4b} \label{P4b} \\
&z_i\in\Kcal_i\cup\{0\}, \quad \forall i\in\Ncal. \tag{P4c} \label{P4c} 
\end{alignat*}
In P4, $\z=(z_i)_{i\in\Ncal}$ are the device decisions for APs. It is possible for multiple APs to serve the same device, i.e., $z_i=z_{i'}$ for $i\neq i'$. Here, $\p = (p_1, \cdots, p_n)$ is the vector consisting of the constant PSDs of all APs over the entire spectrum. P4 has $2n$ variables in total. 

P4 is a fractional programming problem studied by Shen and Yu \cite{shen2017fractional,shen2017fractional2}. Next, we are going to apply a similar technique as in \cite{shen2017fractional2}.
First, apply the Lagrangian dual transform to reformulate P4 to P5 by introducing non-negative auxiliary variable $\bm{\gamma}$:
\begin{alignat*}{2}
\mathop{\text{maximize}}\limits_{\z,\p,\bm{\gamma}}\ \ \ \ &\sum\limits_{i\in\Ncal}c_{z_i}\log(1+\gamma_i) - \sum\limits_{i\in\Ncal}c_{z_i}\gamma_i \\
&+\quad \sum\limits_{i\in\Ncal}\frac{c_{z_i}(1+\gamma_i)p_{i}g_{i\rightarrow z_i}}{n_{z_i} + \sum\limits_{l\in\Ncal_{z_i}}p_{l}g_{l\rightarrow z_i}} \tag{P5a} \label{P5a}\\
\text{subject to} \ \ \ \
&0\leq p_{i}\leq P_{\text{max}}, \quad \forall i\in \Ncal \tag{P5b} \label{P5b} \\
&z_i\in\Kcal_i\cup\{0\}, \quad \forall i\in\Ncal. \tag{P5c} \label{P5c} 
\end{alignat*}
When $\z$ and $\p$ are fixed, we can obtain the optimal $\bm{\gamma}$ by calculating the partial derivative of (\ref{P5a}) with respect to $\bm{\gamma}$:
\begin{align}
\gamma^*_i = \frac{p_i g_{i\rightarrow z_i}}{n_{z_i}+\sum\limits_{l\in\Ncal_{z_i}:l\neq i}p_l g_{l\rightarrow z_i}}. \label{gamma}
\end{align}
P4 and P5 are equivalent because the objective (\ref{P5a}) is concave in $\bm{\gamma}$ and (\ref{P4a}) is equal to (\ref{P5a}) with $\bm{\gamma}^*$ in it.

Next, by applying the quadratic transform technique in \cite{shen2017fractional}, P5 is rewritten in an equivalent form to P6:
\begin{alignat*}{2}
\mathop{\text{maximize}}\limits_{\z,\p,\bm{\gamma},\y}\ \ \ \ &\sum\limits_{i\in\Ncal}c_{z_i}\log(1+\gamma_i) - \sum\limits_{i\in\Ncal}c_{z_i}\gamma_i\\
&+\quad\sum\limits_{i\in\Ncal}\Bigg(2y_i\sqrt{c_{z_i}(1+\gamma_i)p_{i}g_{i\rightarrow z_i}} \\
&-\quad{y_i}^2\Bigg( n_{z_i} + \sum\limits_{l\in\Ncal_{z_i}}p_{l}g_{l\rightarrow z_i}\Bigg)\Bigg) \tag{P6a} \label{P6a}\\
\text{subject to} \ \ \ \
&0\leq p_{i}\leq P_{\text{max}}, \ \ \ \  \forall i\in \Ncal \tag{P6b} \label{P6b} \\
&z_i\in\Kcal_i\cup\{0\}, \ \ \ \  \forall i\in\Ncal \tag{P6c} \label{P6c} 
\end{alignat*}
where $\y$ consists of an array of auxiliary variables. Since (\ref{P6a}) is a quadratic function of each $y_i$, when $\z$, $\p$, and $\bm{\gamma}$ are fixed, we can derive the optimal $\y$ as:
\begin{align}
y^*_i = \frac{\sqrt{c_{z_i}(1+\gamma_i)p_ig_{i\rightarrow z_i}}}{n_{z_i}+\sum\limits_{l\in\Ncal_{z_i}} p_lg_{l\rightarrow z_i}}. \label{y}
\end{align}
Similarly, the optimal power allocation $\p$ for fixed $\z,\y,\bm{\gamma}$ are
\begin{align}
p^*_i = \min \left\lbrace P_{\text{max}}, \frac{c_{z_i}(1+\gamma_i)g_{i\rightarrow z_i}{y_i}^2}{(\sum\limits_{l\in\Ncal_{z_i}}{y_l}^2g_{i\rightarrow s_l})^2} \right\rbrace.  \label{p}
\end{align}
Finally, since the the utility function (\ref{P6a}) can be decoupled for the APs, the optimal $z_i$ should point to the device that maximizes the utility for AP $i$, i.e., 
\begin{align}
z_i=\left\{
\begin{array}{lcl}
0,             ~~ \text{if}~ \mathop{\max}\limits_{j\in\Kcal_i}\{c_j \log(1+\gamma_i) - c_j\gamma_i + \\
2y_i\sqrt{c_j(1+\gamma_i)p_ig_{i\rightarrow j}} - {y_i}^2(n_j+\sum\limits_{l\in\Ncal_j}p_lg_{l\rightarrow j})\}<0\\
\arg \mathop{\max}\limits_{j\in\Kcal_i}\{c_j \log(1+\gamma_i) - c_j\gamma_i + \\
2y_i\sqrt{c_j(1+\gamma_i)p_ig_{i\rightarrow j}} - {y_i}^2(n_j+\sum\limits_{l\in\Ncal_j}p_lg_{l\rightarrow j})\},       \\
\text{otherwise}.
\end{array} \right.  \label{s}
\end{align}
In the case when multiple devices provide the same amount of contribution to the utility function for an AP, we choose the $j$ with minimal index to break the tie.

So far, we derive the optimal $\bm{\gamma}$, $\y$, $\p$, and $\z$ when other variables are fixed. By updating these variables iteratively, it can be easily verified that the utility (\ref{P4a}) is monotonically nondecreasing and hence must converge. This gives us an efficient algorithm referred to as Algorithm~\ref{alg1} to solve P4.

\begin{algorithm}
	\caption{Joint power control and user association for affine utility function}
	\label{alg1}
	\begin{algorithmic}
		\STATE \textbf{Initialization:}
		Pick random $z_i\in\Kcal_i\cup\{0\}$ and $p_i\in[0,P_{\text{max}}]$ for all $i\in\Ncal$.
		\REPEAT
		\STATE Step 1. Update $\bm{\gamma}$ by (\ref{gamma});
		\STATE Step 2. Update $\y$ by (\ref{y});
		\STATE Step 3. Update $\p$ by (\ref{p}); 
		\STATE Step 4. Update $\z$ by (\ref{s});
		\UNTIL Convergence
	\end{algorithmic}
\end{algorithm}	
Algorithm~\ref{alg1} starts with an arbitrary initial power profile $(\z, \p, \bm{\gamma})$ and searches for a single optimal power profile, where APs are shut down sequentially to yield an active subset of APs eventually. The total number of variables needed to be solved at each iteration is $O(n)$, moreover, each iteration in Algorithm~\ref{alg1} is performed in closed form using local variables. In the next section, we are going to combine all the aforementioned propositions and techniques to solve P0 for more general utility functions.

\section{A Scalable Joint Power Allocation and User Association Algorithm}
\label{sec:alg}
Theorem 1 ensures that an optimal allocation partitions the spectrum into no more than $k$ segments. The problem is to identify this set of segments and optimize these bandwidths. Each segment corresponds to a power profile.

We next describe Algorithm~\ref{a:pursuit}, a profile pursuit algorithm which iteratively identifies segment one by one until the utility converges. To initialize, we assume one feasible segment $\p^0$ is given and is added to the set of collected segment $P$. One example of the initial segment can be the conventional full-spectrum-reuse scheme in which each AP reuses the entire spectrum with full transmit power. Denote the resulting rate vector by $\r^0$. With $P$ and $\r^0$ as the input to Algorithm \ref{a:pursuit}, during the $t$-th iteration, one new segment $\p^{t+1}$ is identified by optimizing the local linearization of the utility function at the point $\r^t$ using Algorithm \ref{alg1}. The new segment then is added to the set $P$ if it is not already included, otherwise a randomly generated new segment will be added instead. Then P1 is re-optimized using all the collected segments in $P$ so far to allocate the optimal bandwidth of each segment. Since the set of collected segments grows after each iteration, the utility function increases after each iteration until it converges.

\begin{algorithm}
	\caption{Iterative algorithm for profile pursuit}
	\label{a:pursuit}
	\begin{algorithmic}
		\STATE \textbf{Initialization:}
		Pick full spectrum reuse power allocation denoted by $\p^0$ as the initial feasible solution which gives a feasible rate vector $\r^0$; 
		$t\leftarrow 0$;
		$\Pcal\leftarrow \{\p^0\}$;
		\REPEAT
		\STATE Step 1. Solve P1 with the objective function being $\langle\nabla u(\r^{(t)}), \r\rangle$ using Algorithm \ref{alg1}, which gives us $\p^{t+1}$.
		\STATE Step 2. If $\p^{t+1} \notin \Pcal$, $\Pcal \leftarrow \Pcal\cup\{\p^{t+1}\}$; otherwise, add an arbitrary new segment that is not in $\Pcal$, $t\leftarrow t+1$. 
		\STATE Step 3. Solve P1 given power allocations in $\Pcal$ and obtain the rate vector $\r^{(t)}$.
		\UNTIL Convergence
	\end{algorithmic}
\end{algorithm}	

Algorithm~\ref{a:pursuit} is a Frank-Wolfe type algorithm (also known as the conditional gradient algorithm) \cite{Alg:Frank}. Unique to Algorithm~\ref{a:pursuit} is that each iteration identifies one new power profile to add to the set $\Pcal$ so as to re-optimize P1.

\section{General Resource Allocation Problem}
\label{sec:general}
In this section, the problem formulation of resource management is further generalized to accommodate a much broader set of utility functions, spectral efficiencies, and power constraints. The general problem can also be efficiently solved using the proposed technique in this paper.

First of all, for the spectral efficiency function $s_{i\rightarrow j}(\cdot)$ in (\ref{eq3}), other than using the Shannon formula, it can be any measurable function.

Secondly, the PSD constraint can be more general. For example, the SISO PSD constraint (\ref{eq1}) can be generalized to accommodate multicasting over the same spectrum, i.e.,
\begin{align}
\sum\limits_{j\in\Kcal}{|p_{i\rightarrow j}(f)|}_0 \leq d_i, \ \ \ & \forall i\in \Ncal, f\in[0,W]\label{general:p1}
\end{align}
where $d_i>0$ constrains the degrees of freedom of AP $i$. In an extreme case, $d_i=k$ for all $i\in\Ncal$. That is, an AP may serve all devices using the same slice of spectrum simultaneously. The additional degrees of freedom are particularly useful in the case of multiple-input multiple-output systems with multiple antennas. Moreover, the PSD constraint (\ref{eee}) can also be generalized to the following:
\begin{align}
&0\leq p_{i\rightarrow j}(f)\leq Q, \ \ \ \forall i\in \Ncal, j\in\Kcal, f\in[0,W]. \label{eqnew2}\\ 
&\left\Vert\sum_{j\in\Kcal}p_{i\rightarrow j}\right\Vert_{\alpha}\leq P_{\text{max}}, \ \ \ \forall i\in \Ncal,
\label{eq2}
\end{align} 
where $\left\Vert h\right\Vert_{\alpha} = \left(\int_{0}^{W}{\left(h(x)\right)}^\alpha \dif x\right)^{\frac{1}{\alpha}}$ and $\alpha\geq 1$. The left hand side of (\ref{eq2}) is the $\alpha$-norm of the PSDs allocated to AP $i$. If $\alpha = 1$, (\ref{eq2}) constrains the total power of every AP over the entire available spectrum. On the other hand, when $\alpha\rightarrow \infty$, (\ref{eq2}) becomes the previous peak PSD constraint.

Furthermore, in addition to optimizing throughput and packet delay, we can also incorporate energy efficiency into the utility function. Specifically, APs' transmit powers can be expressed as
\begin{align}
P_i=\sum_{j\in\Kcal}\int_{0}^{W}p_{i\rightarrow j}(f) \dif f,\ \ \ \ i\in\Ncal.
\end{align}
In addition, if AP $i$ can be switched off to save its maintenance power (denoted as $C_i$) when it is not allocated power (i.e., $P_i = 0$), then the utility function regarding energy efficiency can be expressed as minus the total energy cost:
\begin{align}
u_e(\P) = -p\sum_{i\in\Ncal}(P_i+\mathbb{1}_{\{P_i>0\}}C_i),
\end{align}
where $p$ is the ``price'' of power and $\P=(P_i)_{i\in\Ncal}$. Denote the utility function regarding spectral efficiency as $u_s(\r)$, then the utility function that incorporates both spectrum and energy efficiencies is expressed as \\
\begin{align}
U(\r,\P) = u_s(\r) + u_e(\P).
\end{align}

\textit{Theorem 2}:
Let $s_{i\rightarrow j}(\cdot):\mathbb{R}^{n\times k}\rightarrow \mathbb{R}$ be a bounded measurable function. A $(k+n+1)$-piecewise constant power allocation optimally solves the following problem:
\begin{alignat*}{2}
&\mathop{\text{maximize}}\limits_{(r_j), (p_{i\rightarrow j}(f)), (P_i)}\ \ \ \ U(\bm{r},\P)\tag{P7a} \label{P0a}\\
&\text{subject to} \ \ \ \ \\
&r_{j}=\sum\limits_{i\in\Ncal}\int_{0}^{W} s_{i\rightarrow j}(\p(f)) \dif f, \tag{P7b} \quad \forall j\in \Kcal \label{P0b}\\
&P_i=\sum_{j\in\Kcal}\int_{0}^{W}p_{i\rightarrow j}(f) \dif f,\tag{P7c} \quad \forall i\in \Ncal \label{P0c}\\
&\sum\limits_{j\in\Kcal}{|p_{i\rightarrow j}(f)|}_0 \leq d_i, \quad \forall i\in \Ncal, f\in[0,W]\tag{P7d}\label{P0dd}\\
&\left(\int_{0}^{W}\left[\sum_{j\in\Kcal}p_{i\rightarrow j}(f)\right]^{\alpha}\dif f\right)^{\frac{1}{\alpha}}\leq P_{\text{max}}, \quad \forall i\in \Ncal \tag{P7e} \label{P0e} \\
&0\leq p_{i\rightarrow j}(f)\leq Q, \quad \forall i\in \Ncal, j\in\Kcal, f\in[0,W]. \tag{P7f} \label{P0f} 
\end{alignat*}

Theorem 2 is proved in Appendix~\ref{append_A}. P7 describes a very general resource management problem, which essentially encompasses most existing optimization-oriented formulations of physical resource allocation (including spectrum and time slot allocation, user association, and power control) in the literature. The significance here is that the sparsity guaranteed by Theorem 2 allows us to reduce the continuous PSD function allocation problem to that of finding a discrete set of $(k+n+1)$ power profiles.

If the spectral efficiency function $s_{i\rightarrow j}(\cdot)$ can be written as a concave function of the SINR given by (\ref{eqsinr}), that is
\begin{align}
s_{i\rightarrow j}(\p) = A(\gamma_{i\rightarrow j}(\p)),
\label{general:s}
\end{align}
and $A(\cdot): \mathbb{R}\rightarrow\mathbb{R}$ is a concave function, then P7 can be solved using Algorithm~\ref{a:pursuit} proposed in Sec.~\ref{sec:alg}.

\section{Numerical Results}
\label{numerical}
\begin{table}[H]
	\renewcommand{\arraystretch}{1}
	\caption{Parameter Configurations.}
	\label{table_example}
	\centering
	\begin{tabular}{|c|c|c|}
		\hline
		Parameters & Value/Function \\
		\hline
		AP transmit power  & 23 dBm\\
		Total bandwidth &  10 MHz\\
		Average packet length & 0.5 Mbits\\
		AP to device pathloss (LOS) & $30.18+26.7\log_{10}(R)$\\
		AP to device pathloss (NLOS) & $34.53+36\log_{10}(R)$\\
		Shadowing standard deviation (LOS) & 4 dB\\
		Shadowing standard deviation (NLOS) & 10 dB\\
		\hline
	\end{tabular}
\end{table}

In this section, we investigate the performance gain of the proposed allocation scheme by comparing with the following baseline schemes:
\begin{enumerate}
	\item Full-spectrum-reuse $+$ maximum reference signal receive power (MaxRSRP): Every AP reuses all available spectrum with full transmit power and {every} device is associated to the strongest AP in terms of the received power.
	\item Full-spectrum-reuse $+$ optimal user association: {Every} AP reuses all {available} spectrum with full transmit power and {device} association is optimized {for the utility}.
	\item The pattern-pursuit algorithm proposed in \cite{zhou2017cell} which optimizes both spectrum allocation and user association assuming full transmit power at each AP.
\end{enumerate}
The performance metric is the average packet delay which is also the objective in P0. We use parameters compliant with the LTE standard \cite{3GPP:LTE} as given in Table \ref{table_example}. To make fair comparisons, we use exactly the same network (same topology, same channel conditions, and same traffic loads) as in our previous paper \cite{zhou2017cell}, in which all active transmissions use the maximum allowed PSD with binary (0 or peak) power control.
\begin{figure}[t]
	\centering
	\includegraphics[width=3in]{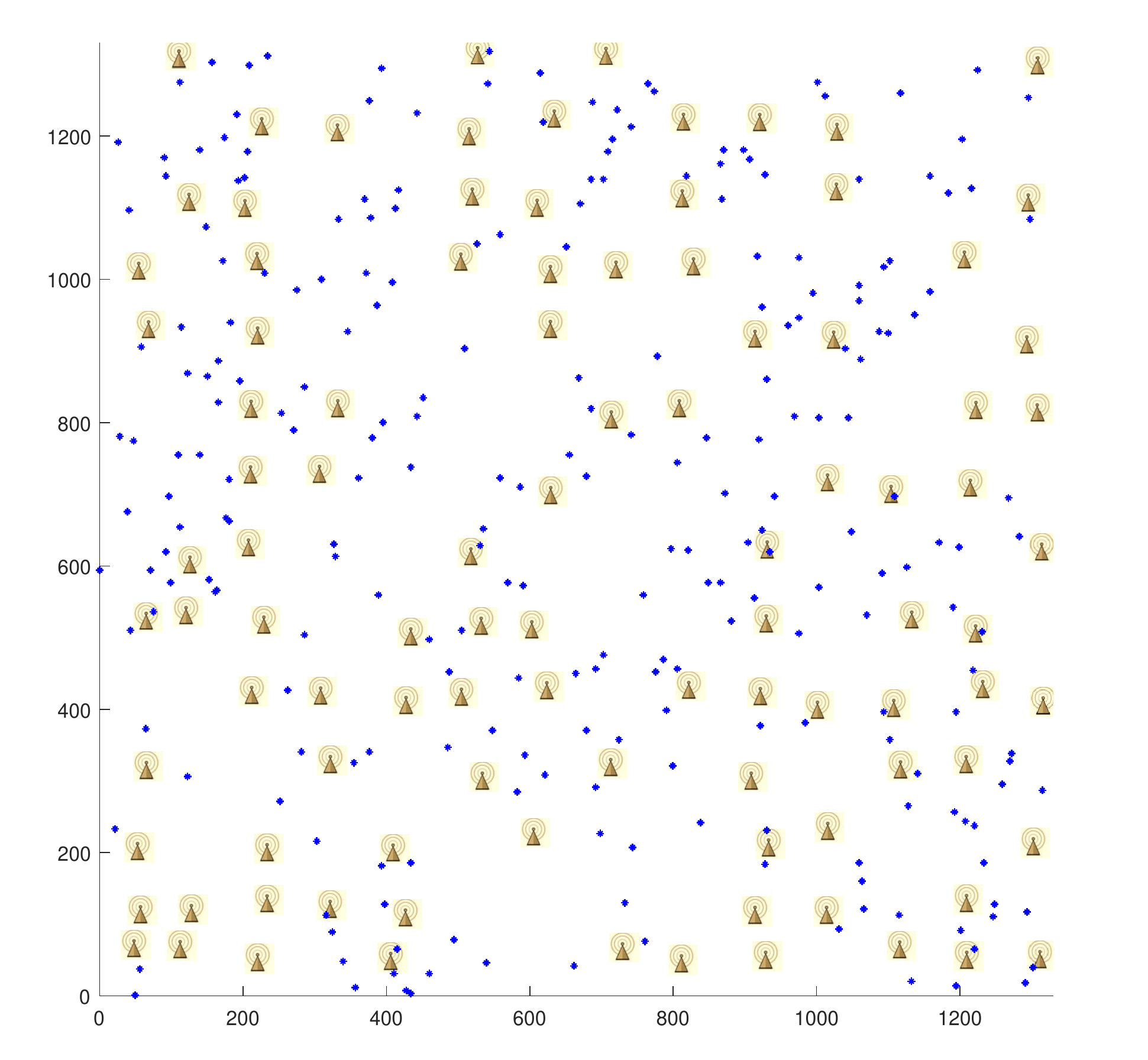}
	\caption{Deployment of the medium scale network.}
	\label{small_deployment}
\end{figure}

\begin{figure}[t]
	\centering
	\includegraphics[width=3.5in]{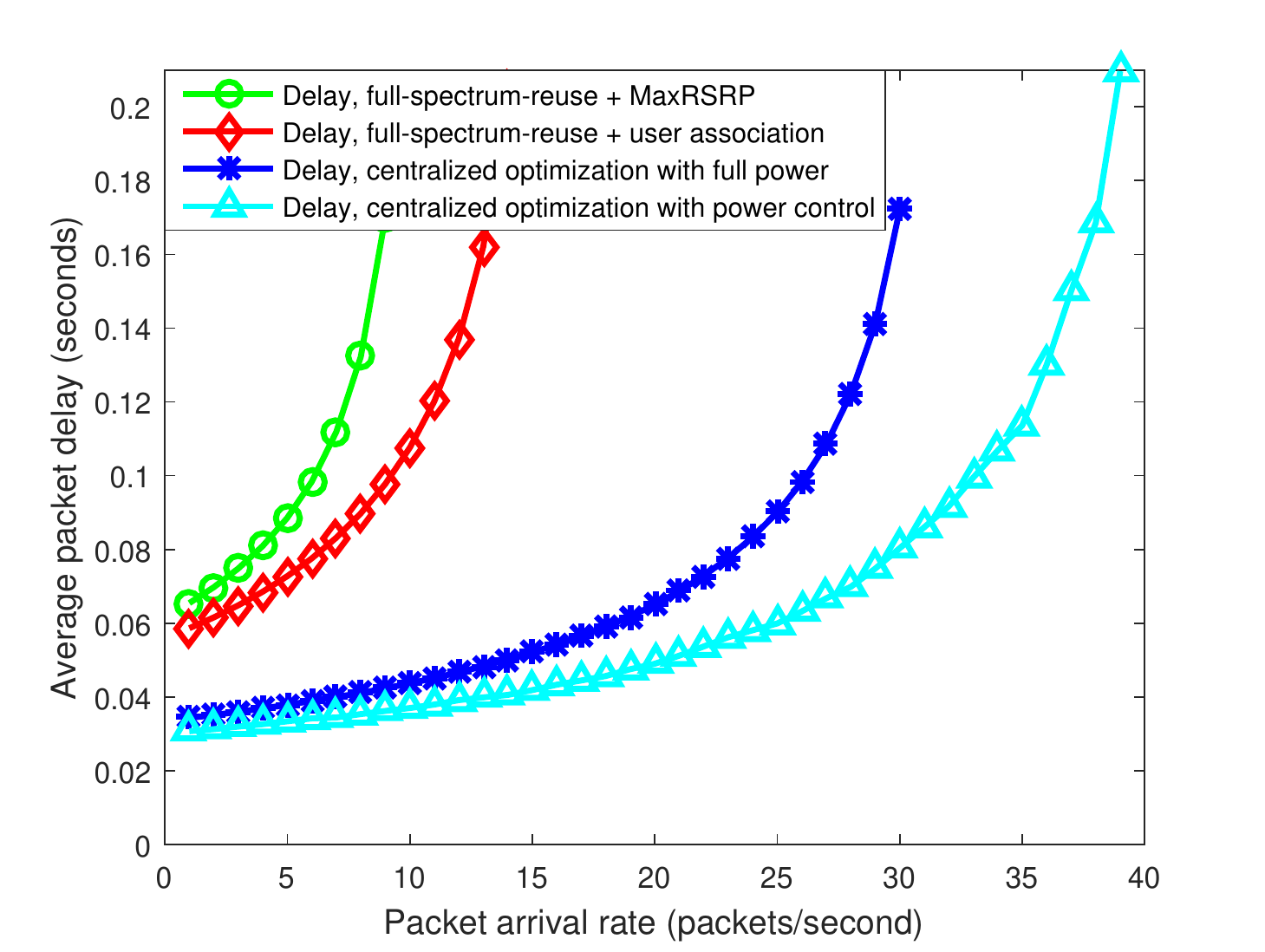}
	\caption{The theoretical packet delays (the objective function) of the proposed scheme and baseline schemes.}
	\label{small_compare}
\end{figure}

\begin{figure}[t]
	\centering
	\includegraphics[width=3.5in]{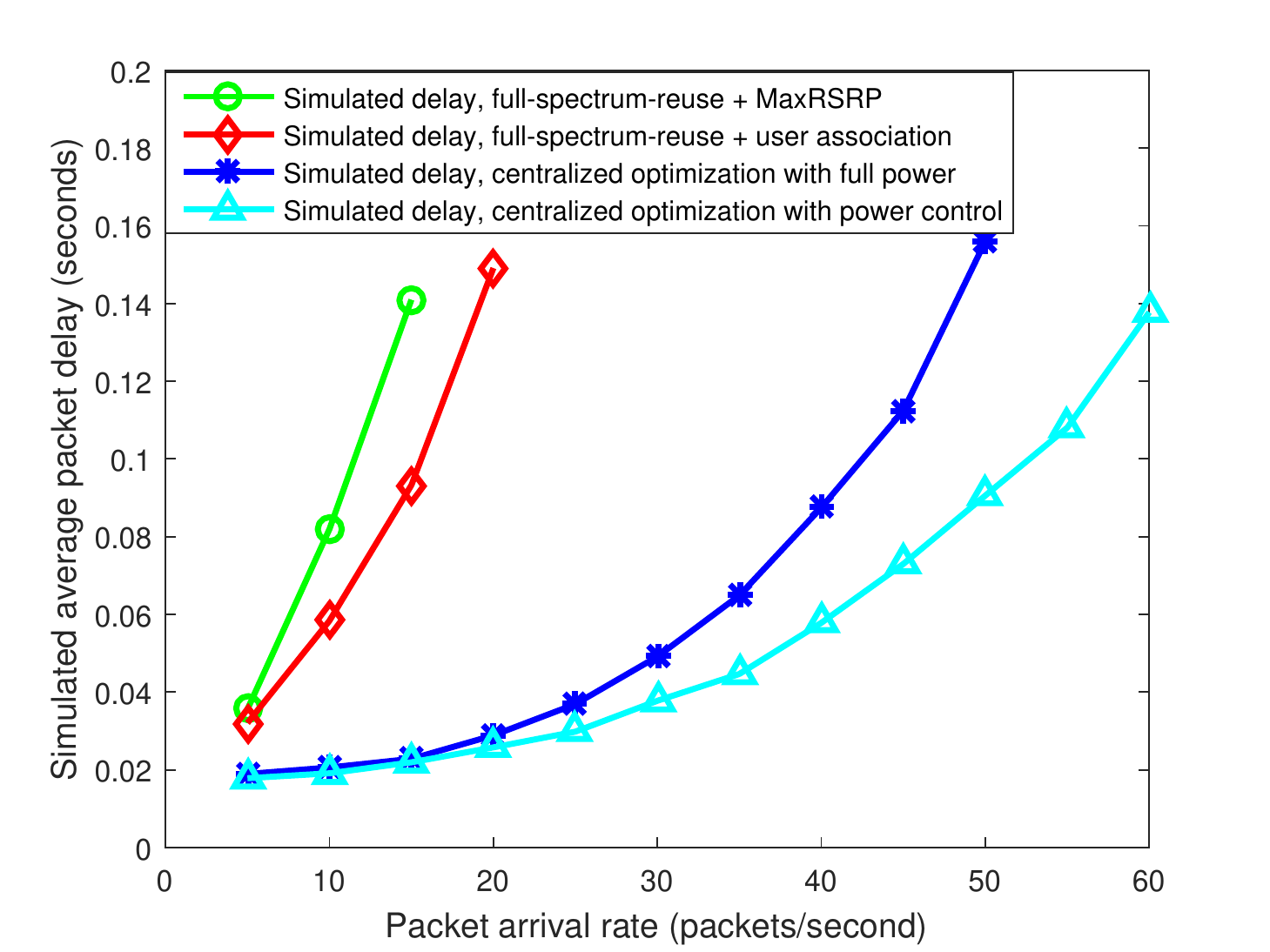}
	\caption{The simulated packet delays of the proposed scheme	and baseline schemes.}
	\label{small_actual}
\end{figure}
\subsection{Performance in Medium Scale Networks}
We first compare the performance of the proposed scheme in a network of medium size. We randomly drop 100 APs and 250 devices over a 1,330 $\times$ 1,330 $\text{m}^2$ area as shown in Fig.~\ref{small_deployment}. The objective (average packet delay) versus traffic intensity curves are shown in Fig.~\ref{small_compare}.

From Fig.~\ref{small_compare}, it can be observed that as the average device traffic increases to above 8, 13, and 30 packets/second, respectively, all three baseline schemes fail to support all the devices. While the proposed solution has significantly larger throughput (above 39 packets/second) than the other schemes. The proposed solution also significantly reduces the delay especially in the high traffic regime. 

To validate the obtained allocation solution, we have built an event-driven packet-level simulator. In the simulator, the instantaneous transmission rate of a link depends on the current set of busy APs occupying the same part of the spectrum. Fig.~\ref{small_actual} compares the simulated average packet delay of the proposed allocation scheme with the baseline schemes. Compared with the theoretical results in Fig.~\ref{small_compare}, all the schemes achieve larger throughput regions. That is because the service rate model (\ref{eq3}) is {``conservative''}, i.e., an AP's transmission rate over any spectrum segment is the worst-case rate, which is the achievable rate when all APs in this segment are transmitting. In addition, the proposed scheme achieves a quite larger throughput (60 packets/second/device) than the  other schemes. Moreover, the delay is also reduced by more than 50\% in the high traffic regime.

\begin{figure}[t]
	\centering
	\includegraphics[width=3.5in]{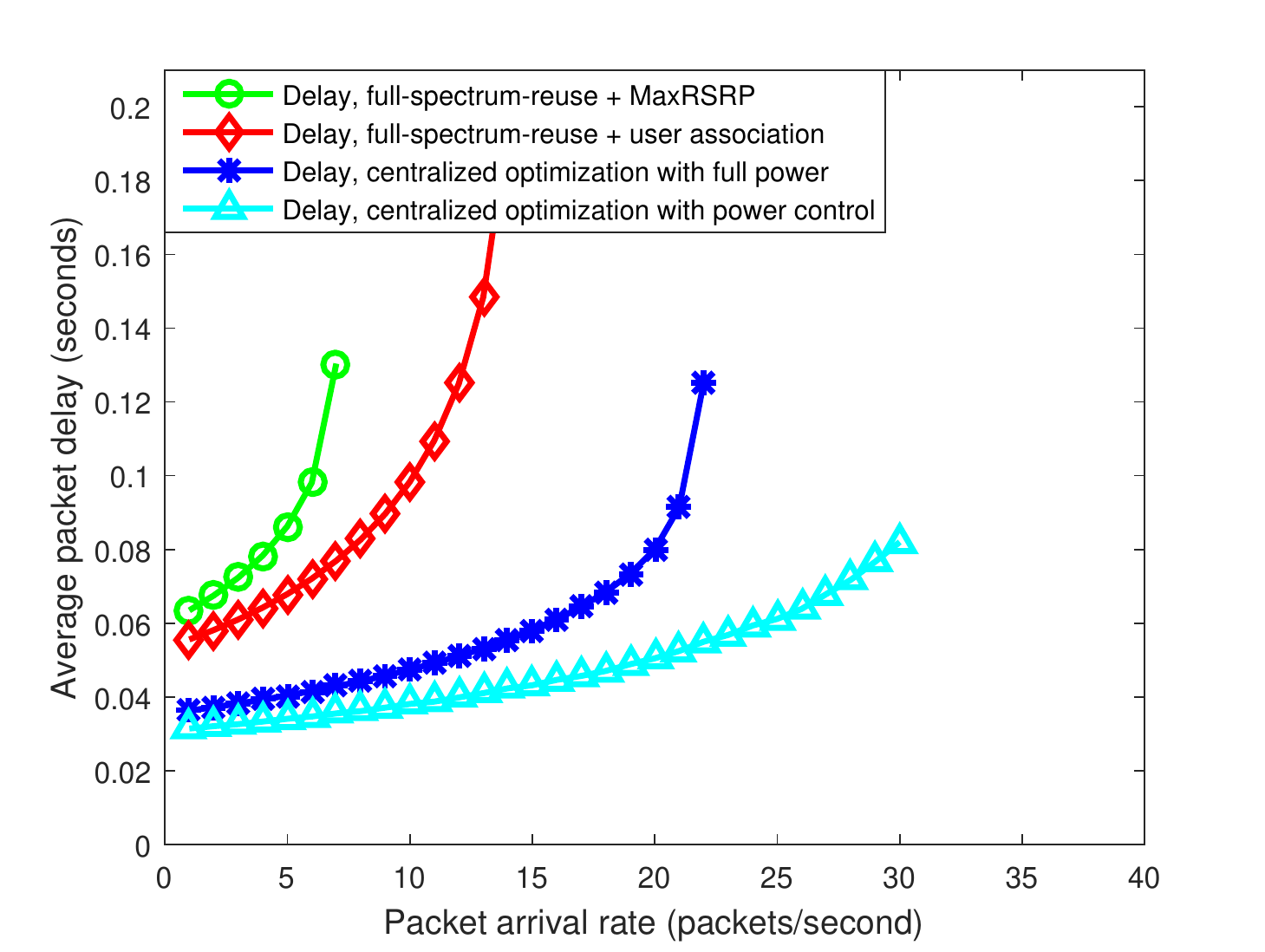}
	\caption{Performance comparison with the baseline schemes for 1,000 APs and 2,500 devices.}
	\label{large_compare}
\end{figure}	

\subsection{Performance in Large Scale Networks}
\begin{figure*}
	\centering
	\subfigure[]{
		\begin{minipage}{9cm}
			\centering
			\includegraphics[width=3in]{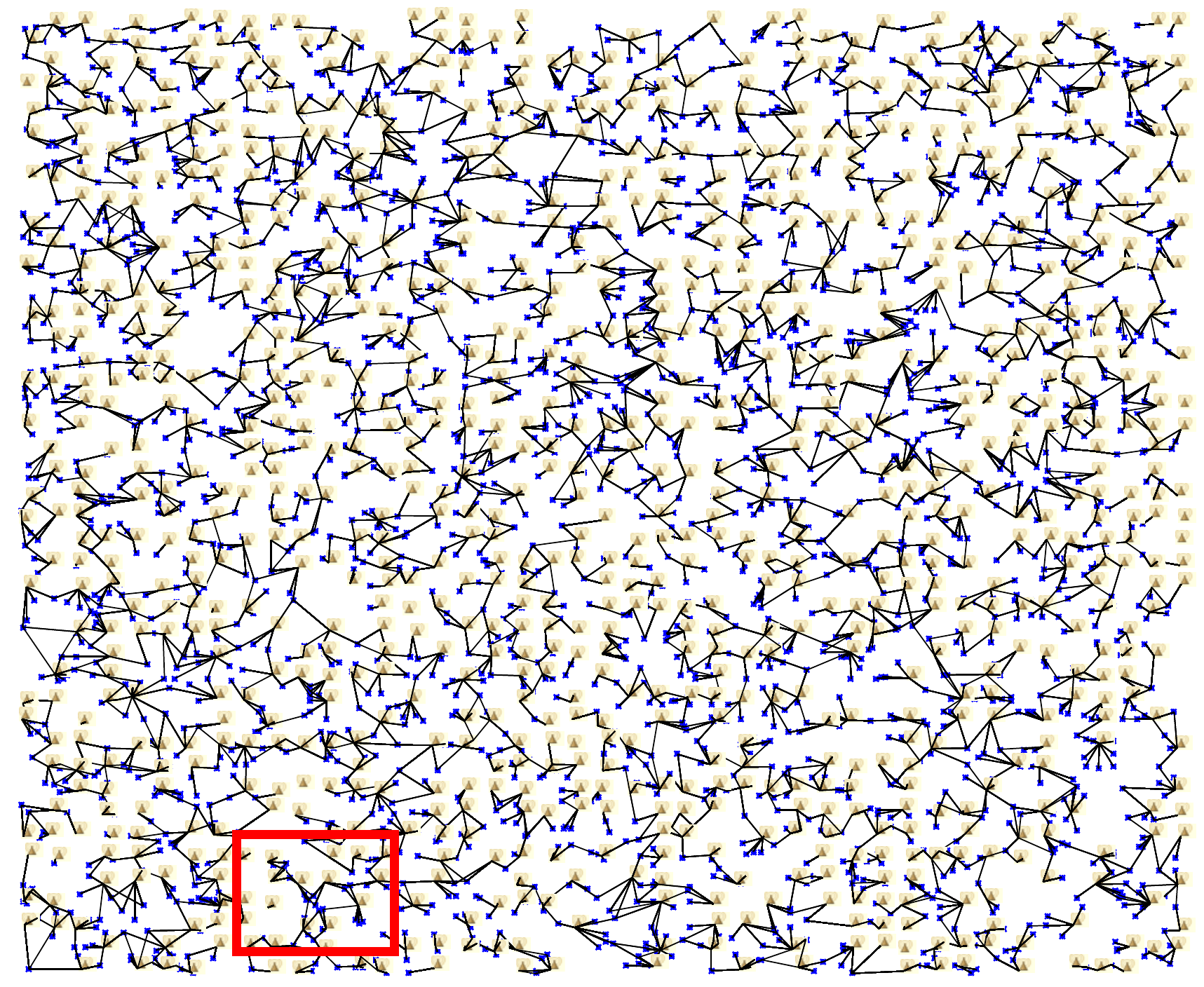}
			\label{zoom_out}
		\end{minipage}
	}
	\subfigure[]{
		\begin{minipage}{7.5cm}
			\centering
			\includegraphics[width=3in]{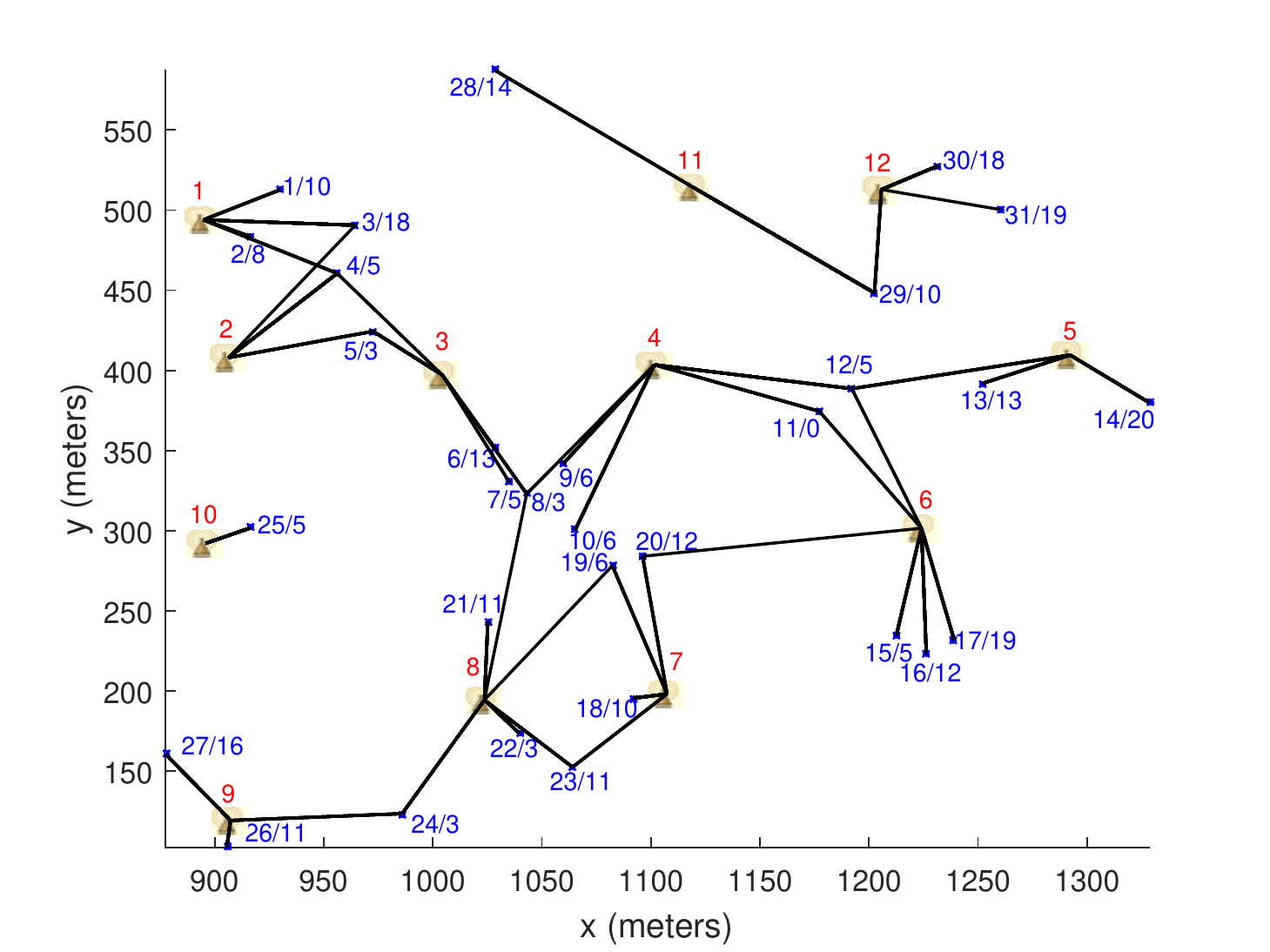}
			\label{zoom_in}
		\end{minipage}
	}
	\subfigure[]{
		\centering
		\includegraphics[width=6in]{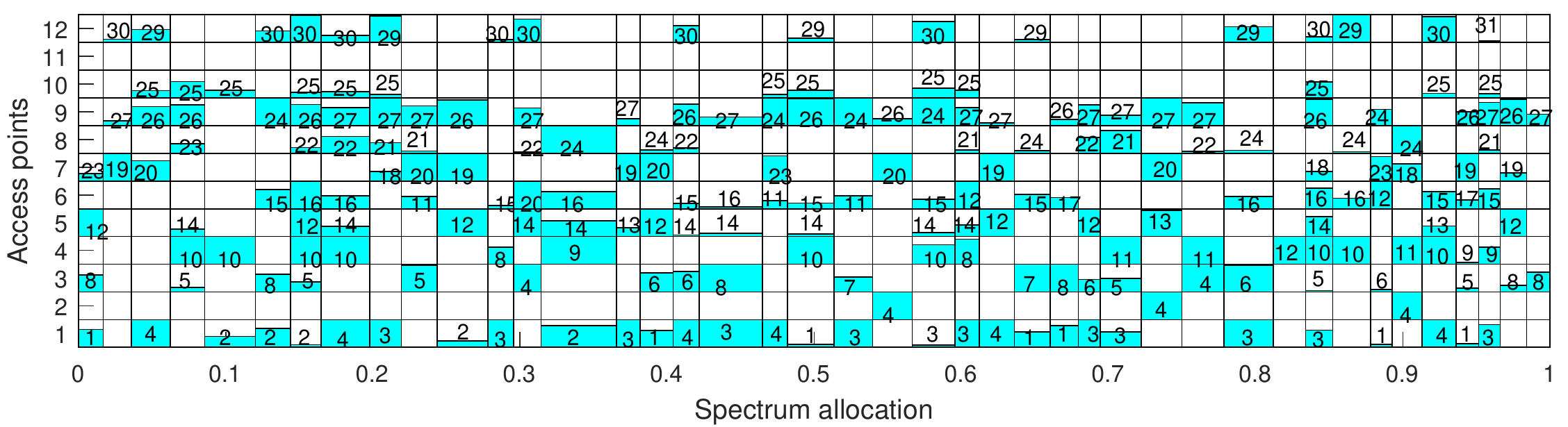}
		\label{all}
	}
	\caption{Resource management in very large networks. (a) Deployment and user association for the large network. (b) Topology graph for the marked area in Fig.~\ref{zoom_out}. (c) Allocation graph for the marked area in Fig.~\ref{zoom_out}.} \label{figall}
\end{figure*}
In this subsection, the proposed allocation scheme is used to compute the allocation solution for a large network of 1,000 APs and 2,500 devices over a 4,200~m $\times$ 4,200~m area. The AP density remains the same as that of the medium-scale network. The objective (average packet delay) versus traffic intensity curves are shown in Fig.~\ref{large_compare}.

From Fig.~\ref{large_compare}, it is clear that the proposed solution has significantly larger throughput (above 30 packets/second) than full-spectrum reuse with maxRSRP association (7 packets/second), full-spectrum reuse with optimal user association allocation (14 packets/second) and centralized optimization with full power (22 packets/second). The proposed solution also outperforms other schemes in delay especially in the high traffic regime. 

The obtained resource allocation at arrival rate of 25 packets/second/device is shown in Fig.~\ref{figall}. As shown in Fig.~\ref{zoom_out}, the lines connecting each device-AP pair indicate an association. To clearly present spectrum allocation, user association, and power allocation, the local cluster as marked in the red rectangle region in Fig.~\ref{zoom_out} is shown in enlarged display in Figs.~\ref{zoom_in} and \ref{all}. Fig.~\ref{zoom_in} shows the user association for the marked area. The numbers above each device represent the device index and its traffic load, respectively. The number above each AP represents the AP index. The spectrum and power allocation for the marked area is shown in Fig.~\ref{all}.  The widths of the rectangles represent fractions of the entire spectrum of the active segments. The solid ones in each row are the spectrum segments that are used by the corresponding AP to serve the device whose index is marked on that spectrum segment. Therefore, each row can be viewed as the PSD of the corresponding AP. Strongly interfering links are usually assigned different spectrum segments (e.g., link $2\to 4$ and link $3\to 5$), and the same spectrum segments are reused by two links that are far apart (e.g., link $10\to 25$ and link $9\to 26$). Moreover, devices with light traffic loads or devices on the transmission edge of two APs (e.g., device 5) are assigned less spectrum and less power, and vice versa. These numerical results demonstrate that the algorithm achieves topology aware frequency reuse for interference management, as well as an efficient traffic aware spectrum allocation.

\subsection{Computational Complexity}
Table \ref{runtime} shows the computational cost for the proposed method under different network scales. The density of APs and devices remains the same for all networks. One observation is that the number of iterations for Algorithm \ref{alg1} to converge varies slightly even when the network size is enlarged by hundred-fold. The algorithms are run on an Intel Core Xeon 3.6 GHz 6-core computer. It can be observed that the total computation only takes 1.8 seconds for small-scale networks and 170 seconds for metropolitan-scale networks because of the closed-form updates in every iteration.\footnote{The runtime is expected to be reduced significantly to fit in a moderate timescale using a powerful cluster/cloud.} In addition, the number of active segments is very close to the number of candidate segments for all networks, meaning that at each iteration of Algorithm \ref{a:pursuit}, a useful segment is almost always obtained to contribute to the final solution.

The proposed algorithm is able to solve the resource allocation problem for super large networks using small amount of time due to its small number of iterations and low per-iteration cost. As only convergence to a local optimum is guaranteed in all cases, the converged values may differ depending on the starting point.

\begin{table}[t]
	\renewcommand{\arraystretch}{1.3}
	\caption{Computational cost for Algorithm \ref{alg1} and Algorithm \ref{a:pursuit}}
	\label{runtime}
	\centering

		\begin{tabular}{||c|c|c|c||}
			\hline
			\hline
			Networks scales & 10 APs  & 100 APs  & 1,000 APs\\
			& 25 devices & 250 devices & 2,500 devices\\
			\hline
			\# of iterations in Alg.~\ref{alg1} & 85.0 & 92.5 & 122.6  \\
			\hline
			\# of iterations in Alg.~\ref{a:pursuit} & 12.8 & 23.8 & 27.2 \\
			\hline
			\# of active segments & 10.6 & 23.1 & 27.2\\
			\hline
			Average runtime (seconds) &1.8 &9.5  &169.9\\ 
			\hline
	\end{tabular}
\end{table}

\section{Conclusion}
\label{conclusion}
We have studied a challenging joint resource allocation problem in metropolitan-scale networks. We have developed a scalable formulation by exploiting the hidden sparse structure of an optimal allocation and proposed an efficient algorithm to solve the reformulated problem with guaranteed convergence. Moreover, each iteration is performed in closed form, which makes centralized resource allocation practically feasible even for very large networks. From numerical results, significant gains are observed compared with conventional resource management schemes.\\

One possible future work is to incorporate the slow-timescale solution with fast-timescale link scheduling \cite{teng2015Dual}. That is, the slow-timescale solution may guide each AP to better schedule transmission links locally on a fast-timescale. Another extension is to consider allocating discrete subcarriers rather than continuous spectrum resources, which may involve more complex problem formulations. This will allow the proposed solution to be adopted in current and emerging radio access network (RAN) standards.

\section*{Acknowledgment}
The authors thank Weimin Xiao, Jialing Liu, Michael Honig, and Ermin Wei for stimulating discussions.


%

\appendices
\section{Proof of Theorem 2}
\label{append_A}
The PSDs in P7 are arbitrary (Lebesgue measurable) functions in an infinite dimensional function space. They determine the utility function through the rate and power vectors ($\r$ and $\P$) of finite dimensions. The key to the proof is to invoke the Carath\'eodory's theorem \cite{Egg:Conv} to assert the existence of a $(k+n+1)$-dimensional allocation that achieves the desired $(\r,\P)$.

Let $\mathcal{L}$ denote the set of Lebesgue integrable functions on $[0,W]$. Let us define:
\begin{align}
\begin{split}
\Rcal = \Bigg\{ &(\r, \P, \t)\in \mathbb{R}^{k}\times\mathbb{R}^{n}\times\mathbb{R}^{n}: \exists~ \p(\cdot)=(p_{i\rightarrow j}(\cdot))\in{\mathcal{L}}^{nk}\\
&\text{s.t.}~~r_{j}=\sum\limits_{i\in\Ncal}\int_{0}^{W} s_{i\rightarrow j}(\p(f)) \dif f,~~\forall j\in\Kcal,\\
&P_i=\sum_{j\in\Kcal}\int_{0}^{W}p_{i\rightarrow j}(f) \dif f,~~\forall i\in\Ncal,\\
&t_i=\int_{0}^{W}\left[\sum\limits_{j\in\Kcal}p_{i\rightarrow j}(f)\right]^{\alpha} \dif f,~~\forall i\in\Ncal,\\
&t_i\leq (P_{\text{max}})^{\alpha},~~\forall i\in\Ncal,\\
&\sum\limits_{j\in\Kcal}{|p_{i\rightarrow j}(f)|}_0 \leq d_i,~~\forall i\in\Ncal, f\in[0,W],\\
&0\leq p_{i\rightarrow j}(f)\leq Q,~~\forall i\in\Ncal,j\in\Kcal,f\in[0,W]  \vphantom{(\r, \P, \t)} \Bigg\}.
\end{split}
\label{region}
\end{align}
We adopt the Lebesgue integral throughout as a convenient justification of using ``maximize'' in P0 and P7, i.e., that the maximum is actually achieved. While it is perhaps possible to adopt the less technical Riemann integral to the same outcome, it is not as easy to work with. In particular, the limit of a sequence of Riemann integrable functions may not be Riemann integrable at all.

In the special case P0, as long as $s_{i\rightarrow j}(\p(\cdot))$ is measurable, $\p$ does not have to be measurable. Nonetheless, we can restrict to measurable ones without loss of generality because any feasible rate vector can be constructed by a measurable $\p$.

We can think of $\Rcal$ as a manifold in the $(k+2n)$-dimensional Euclidean space induced by (continuous) power allocations $(p_{i\rightarrow j}(f))_{i\in\Ncal,j\in\Kcal,f\in[0,W]}$ that satisfy some power constraints. $\Rcal$ is nonempty because it includes the vector induced by the all-zero power allocation. For every $(\r, \P, \t)\in\Rcal$, $(\r,\P)$ are a pair of feasible rate and power vectors. The reason to also include the vector $\t$ becomes clear later.

Define another set
\begin{align}
\begin{split}
\Scal = \Bigg\{ &(\u, \v, \w)\in\mathbb{R}^{k}\times\mathbb{R}^{n}\times\mathbb{R}^{n}:\exists~\q\in{[0,Q]}^{nk}\\
&\text{s.t.}~~{u}_j=W \sum_{i\in\Ncal}s_{i\rightarrow j}(\q),~~\forall j\in\Kcal,\\
&v_i=W\sum_{j\in\Kcal}q_{i\rightarrow j},~~\forall i\in\Ncal,\\
&w_i=W\left[\sum\limits_{j\in\Kcal}q_{i\rightarrow j}\right]^{\alpha},~~\forall i\in\Ncal,\\
&\sum\limits_{j\in\Kcal}{|q_{i\rightarrow j}|}_0 \leq d_i, \forall i\in\Ncal \Bigg\}.
\end{split} 
\label{ins_link}
\end{align}
The set $\Scal$ is bounded, closed, and therefore compact. $\Scal$ is also a manifold in the $(k+2n)$-dimensional space. Every vector in $\Scal$ is induced by a single power profile $\q\in{[0,Q]}^{nk}$, which can be achieved by the flat power allocation over the entire bandwidth defined according to $p_{i\rightarrow j}(f)=q_{i\rightarrow j},~\forall i\in\Ncal,j\in\Kcal,f\in[0,W]$. If indeed this single power profile is adopted over $[0,W]$, we have according to (\ref{ins_link}):
\begin{align}
\begin{split}
&r_{j}=\sum\limits_{i\in\Ncal}\int_{0}^{W} s_{i\rightarrow j}(\p(f)) \dif f=W\sum_{i\in\Ncal}s_{i\rightarrow j}(\q)=u_j,\\
&P_i=\sum_{j\in\Kcal}\int_{0}^{W}p_{i\rightarrow j}(f) \dif f = W\sum_{j\in\Kcal}q_{i\rightarrow j} = v_i,\\
&\text{and}\\
&t_i = \int_{0}^{W}\left[\sum\limits_{j\in\Kcal}p_{i\rightarrow j}(f)\right]^{\alpha} \dif f = W\left[\sum\limits_{j\in\Kcal}q_{i\rightarrow j}\right]^{\alpha} = w_i\\
&\text{for all}~i\in\Ncal~\text{and}~j\in\Kcal.
\end{split}
\label{eq222}
\end{align}

The remainder of the proof includes three technical steps: We first show that $\Rcal$ is a subset of the convex hull of $\Scal$, i.e., $\Rcal\subset\text{conv}(\Scal)$. We then invoke the Carath\'eodory's theorem to show that every point in $\Rcal$ is a convex combination of $(k+2n+1)$ points in $\Scal$, and is hence achieved using a $(k+2n+1)$-piecewise constant power allocation. Finally, we reduce the number of sub-bands needed from $k+2n+1$ to $k+n+1$.

Step 1: The goal of this step is to show $\Rcal\subset\text{conv}(\Scal)$. Because $\Scal$ is compact, $\text{conv}(\Scal)$ must be closed. Given $(\r,\P,\t)\in\Rcal$, if we can explicitly define a sequence $\x^1,\x^2,\cdots\in\text{conv}(\Scal)$, which converges to $(\r,\P,\t)$, then $(\r,\P,\t)\in\text{conv}(\Scal)$. This part of the proof hinges on the simple function approximation of Lebesgue measurable functions.\footnote{This approach does not apply to Riemann integrable functions in general.}

Specifically, for every $(\r, \P, \t)\in\Rcal$, there exists a power allocation $\p(\cdot)$ that lies in the feasible region of P7, which results in the rate vecor $\r$ and tranmist power vector $\P$ as well as $\t$ as defined in (\ref{region}). Since the PSD of each AP is a bounded measurable function over $[0,W]$, it can be arbitrarily closely approximated by a sequence of {\em simple functions} \cite{Terse:John}. Fix a positive integer $l$, and partition the $kn$-dimensional cube $[0,\sqrt{l})^{kn}$ into $l^{kn}$ disjoint cubes $\{C^m\}_{m=1}^{l^{kn}}$ with equal side length $\frac{1}{\sqrt{l}}$ on each dimension. Each dimension of the cube is indexed by a pair $(i,j)$ with $i\in\Ncal$ and $j\in\Kcal$. Let the edge of partition $C^m$ in the $(i,j)$-th dimension be represented by $[a^m_{i\rightarrow j},a^m_{i\rightarrow j}+\frac{1}{\sqrt{l}})$. Then $C^m=\prod_{i\in\Ncal}\prod_{j\in \Kcal}[a^m_{i\rightarrow j},a^m_{i\rightarrow j}+\frac{1}{\sqrt{l}})$. Since $\p$ is a measurable function, we can define measurable partitions of $[0,W]$ by $A^m=\p^{-1}(C^m),~m=1,\cdots,l^{kn}$ and $A^0=[0,W] \backslash \bigcup\limits_{m=1}^{l^{kn}} A_{m}$. $A^m$ describes the spectrum on which the power profiles find their values in partition $C^m$. We define a simple function $\p^l: [0,W]\rightarrow [0,\sqrt{l}]^{kn}$ as
\begin{align}
p^l_{i\rightarrow j}(f)=\sum_{m=1}^{l^{kn}}a^{m}_{i\rightarrow j}\mathbb{1}_{\{f\in A^m\}}.
\end{align}
It is obvious that $p^l_{i\rightarrow j}(f)\leq p_{i\rightarrow j}(f)$ for all $f\in[0,W], i\in\Ncal, j\in\Kcal$. As $l\rightarrow \infty$, this sequence of simple functions $(p^l_{i\rightarrow j})$ converges to $(p_{i\rightarrow j})$ from below almost everywhere. Moreover, let $\r^l=(r^l_j)_{j\in\Kcal}$ where $r^l_j=\sum_{m=1}^{l^{kn}}\sum_{i\in \Ncal}S_{i\rightarrow j}(a^m_{i\rightarrow j})\mu(A^m)$,\\ 
let $\P^l=(P^l_i)_{i\in\Ncal}$ where $P^l_i=\sum_{m=1}^{l^{kn}}\sum_{j\in \Kcal}a^m_{i\rightarrow j}\mu(A^m)$, \\
and let $\t^l = (t^l_i)_{i\in\Ncal}$ where $t^l_i=\sum_{m=1}^{l^{kn}}[\sum_{j\in \Kcal}a^m_{i\rightarrow j}]^{\alpha}\mu(A^m)$,\\
where $\mu(\cdot)$ is the Lebesgue measure in $\mathbb{R}^{kn}$. It is clear that $\r^l\leq\r$ and $\P^l\leq\P$. And we also have 
\begin{align}
t^l_i&=\int_{0}^{W}\left[\sum\limits_{j\in\Kcal}p^l_{i\rightarrow j}(f)\right]^{\alpha} \dif f\\
&\leq\int_{0}^{W}\left[\sum\limits_{j\in\Kcal}p_{i\rightarrow j}(f)\right]^{\alpha} \dif f\\
&\leq (P_{\text{max}})^{\alpha},\quad \forall i\in\Ncal.
\end{align} 
Therefore, $(\r^l, \P^l, \t^l)\in\text{conv}(\Scal)$ and $(\r^l, \P^l, \t^l)\leq(\r, \P, \t)$. As $l$ goes to $\infty$, the number of cubes in which $[0,Q]^{kn}$ is partitioned goes to $\infty$, $(\r^l, \P^l, \t^l)$ converges to $(\r, \P, \t)$. Therefore, $\Rcal\subset\text{conv}(\Scal)$.

Step 2: By Carath\'eodory's theorem \cite{Egg:Conv}, $(\r, \P, \t)$ can be constructed as a convex combination of $k+2n+1$ vectors in $\Scal$.\footnote{Carath\'eodory's theorem has been invoked in the past to establish similar results~(e.g., \cite{Etkin:Spectrum} and \cite{Large:Zhuang}), but this is the first analysis that rigorously examine all conditions.} That is, for any feasible solution $(\r, \P, \t)$ of P7, it can be rewritten as a convex combination of $k+2n+1$ points in $\Scal$. That is 
\begin{align}
(\r, \P, \t) = \sum\limits_{m=1}^{k+2n+1}\beta^m(\u^m, \v^m, \w^m),
\end{align}
where $(\u^m, \v^m, \w^m)\in\Scal$, $\beta^m\geq0$, $m=1,\cdots,k+2n+1$, and $\sum\limits_{m=1}^{k+2n+1}\beta^m = 1$. Specifically, once the $k+2n+1$ power profiles and their weights are known, the power allocation $\p$ can be constructed as
\begin{align}
p_{i\rightarrow j}(f)=q^l_{i\rightarrow j},~~\text{if}~f\in\left[\sum\limits_{m=1}^{l-1}\beta^mW,\sum\limits_{m=1}^{l}\beta^mW\right]\notag\\\text{for}~l=1,\cdots,k+2n+1.
\end{align}
As in (\ref{eq222}), it is straightforward to verify that this $(k+2n+1)$-piecewise constant power allocation is feasible and achieves the optimal utility. Therefore, any feasible solution to P7 can be attained with a $(k+2n+1)$-piecewise constant power allocation.

Step 3: We next reduce the number of sub-bands needed from $k+2n+1$ to $k+n+1$. 

For any feasible solution $(\r, \P)$ of P7, as we just proved, we can rewrite it as a convex combination of $k+2n+1$ points in $\Scal$. That is
\begin{align}
(\r, \P) = \sum\limits_{m=1}^{k+2n+1}\beta^m(\u^m, \v^m),
\end{align}
where $(\u^m, \v^m)\in\Scal$, $m=1,\cdots,k+2n+1$. Moreover, we have
\begin{align}
&\sum\limits_{m=1}^{k+2n+1}\beta^m=1,\\
&\sum\limits_{m=1}^{k+2n+1}\beta^m\w^m\leq (P_{\text{max}})^{\alpha}.
\end{align}

Since $k+2n+1>\text{dim}\left((1, \r, \P)\right)=k+n+1$, row vectors $(1, \u^1, \v^1),\cdots,(1, \u^{k+2n+1}, \v^{k+2n+1})$ must be linearly dependent. Therefore, there are real scalars $\mu^m$, $m=1,\cdots,k+2n+1$, not all zero, such that
\begin{align}
\sum\limits_{m=1}^{k+2n+1}\mu^m=0,\label{eq24}
\end{align}
and
\begin{align}
\sum\limits_{m=1}^{k+2n+1}\mu^m(\u^m, \v^m) = 0.\label{eq25}
\end{align}
Moreover, the coefficients can be chosen to satisfy
\begin{align}
\sum\limits_{m=1}^{k+2n+1}\mu^m\w^m\geq 0. \label{eq26}
\end{align}
because if (\ref{eq26}) does not hold, we can replace $\mu^m$ by $-\mu^m$ to satisfy (\ref{eq24})--(\ref{eq26}). Then $(\r, \P)$ can be rewritten as 
\begin{align}
(\r, \P) = \sum\limits_{m=1}^{k+2n+1}(\beta^m-a\mu^m)(\u^m, \v^m),
\end{align}
for any real-valued $a$. Since not all of the $\mu^m$ are equal to zero. Therefore, there exists at least one $\mu^m>0$. Define
\begin{align}
\hat{a} = \mathop{\text{min}}\limits_{1\leq m\leq k+2n+1}\left\lbrace \frac{\beta^m}{\mu^m}:~\mu^m>0\right\rbrace = \frac{\beta^q}{\mu^q},
\end{align} 
where $q$ is an index with 
\begin{align}
\beta^q-\hat{a}\mu^q=0.
\label{eqeq}
\end{align}
Note that $\hat{a}>0$, and for every $m$ between $1$ and $k+2n+1$,
\begin{align}
\beta^m-\hat{a}\mu^m\geq 0.
\end{align}
Therefore,
\begin{align}
(\r, \P) = \sum\limits_{m=1}^{k+2n+1}(\beta^m-\hat{a}\mu^m)(\u^m, \v^m),
\end{align}
with 
\begin{align}
\sum\limits_{m=1}^{k+2n+1}(\beta^m-\hat{a}\mu^m)\w^m &\leq \sum\limits_{m=1}^{k+2n+1}\beta^m\w^m - 0 \notag\\
&\leq (P_{\text{max}})^{\alpha}.
\end{align}
Every coefficient $\beta^m-\hat{a}\mu^m$ is nonnegative, their sum is one, and furthermore, the index $q$ satisfies (\ref{eqeq}). In other words, $(\r, \P)$ is represented as a convex combination of at most $k+2n$ points of $\Scal$. This process can be repeated until $(\r, \P)$ is represented as a convex combination of at most $k+n+1$ points in $\Scal$. \hfill $\blacksquare$

\section{Proof of Theorem 1}
\label{append_B}
P0 is a special case of P7. Using the same technique developed in Appendix~\ref{append_A} (and dropping the $\P$ dimension in defining $\Rcal$), we can show that the optimal utility of P0 can be achieved by a $(k+1)$-piecewise constant power allocation. Moreover, an optimal solution $\r^*$ must be found on the boundary of the feasible set of rate vectors. (If one has an optimal rate vector being an interior point, one can increase all of its dimensions until reaching the boundary with no loss of utility.) Thus, $k$-piecewise constant power allocation achieves the optimum of P0. \hfill $\blacksquare$


\ifCLASSOPTIONcaptionsoff
  \newpage
\fi

\end{document}